\documentstyle[multicol,aps,prb,graphicx]{revtex} 
\begin{document} 
\draft
\title{Dynamical Mean-Field Theory of Electron-Phonon Interactions in Correlated Systems: Application to Isotope Effects on Electronic Properties} 
\author{Andreas Deppeler and A. J. Millis}
\address{Center for Materials Theory, Department of Physics and Astronomy,
Rutgers University, Piscataway, New Jersey 08854} 
\date{\today}
\maketitle
\begin{abstract}
We use a recently developed formalism (combining an adiabatic expansion and dynamical mean-field theory) to obtain expressions for isotope effects on electronic properties in correlated systems. As an example we calculate the isotope effect on electron effective mass for the Holstein model as a function of electron-phonon interaction strength and doping. Our systematic expansion generates diagrams neglected in previous studies, which turn out to give the dominant contributions. The isotope effect is small unless the system is near a lattice instability. We compare this to experiment.
\end{abstract}
\begin{multicols}{2} 
\section{Introduction}
In solid state physics the term ``isotope effect'' has come to mean a dependence of an electronic property on an ionic mass. The isotope effect on the superconducting transition temperature $T_c$ of conventional superconductors was important evidence for the role played by phonons in the pairing mechanism.\cite{m50,rswn50} In conventional metals the isotope effect on electronic properties other than the superconducting $T_c$ is negligible. The Migdal-Eliashberg (ME) theory of electron-phonon coupling in metals \cite{m58} explains this as follows: most electronic properties are determined by processes occurring on the scale of the electron kinetic energy $t$. Interaction with phonons affects electronic properties only on the scale of a typical phonon frequency $\omega_0$, which is much less than $t$. Thus isotope effects are generically expected to be small, of the order of the adiabatic parameter $\gamma = \omega_0/t \ll 1$. ME theory essentially retains terms only of order $\gamma^0$; therefore, most isotope effects are beyond the scope of this theory. The only exception is the superconducting transition temperature $T_c = \omega_0 \exp(-1/[\lambda - \mu^{\ast}(\omega_0)])$ where the phonon frequency enters as the upper cutoff of the logarithmic divergence in the pairing interaction and the lower cutoff of the logarithmic divergence in the Coulomb pseudopotential $\mu^{\ast}$.

Recent observations in several classes of ``strongly correlated'' materials (high-temperature superconductors, ``colossal magnetoresistance" manganites, and alkali-metal doped ${\rm C}_{60}$) of large isotope effects on electronic properties, including electron effective mass \cite{zm95,zssm95,zhkm97,zckm98,zkc01} and superconducting,\cite{zm95,zssm95,zhkm97,zckm98,zkc01,f99} magnetic,\cite{zckm96,bbgkbok98,ficci98,heil00} and charge ordering\cite{zgg98,if98,betal99,mi99} transition temperatures pose a fundamental challenge to this understanding and call for a theory that goes beyond ME. Some authors \cite{zhkm97,zckm98,zckm96,ficci98} have attempted to relate their experimental data to formulas derived for the case of a ``polaron'': a single electron interacting with lattice deformation. All experimentally relevant systems, however, have a metallic density of electrons, of order one per unit cell, so the applicability of ``polaron'' formulas is not clear. Others \cite{gcp98} have considered the first corrections to ME theory for the electron mass using standard diagrammatic methods but have not considered the feedback effects of electrons on phonons and have presented results that cannot easily be extended to correlated systems.

Very recently, we have introduced a new method,\cite{dm00} combining the dynamical mean-field (DMF) theory\cite{gkkr96} and an adiabatic (small-$\gamma$) expansion, for studying electron-phonon interactions in systems with arbitrary electronic correlations. Here we use this method to calculate the electron effective mass $m^{\ast}$ due to interactions with phonons. Our analytic result includes order-$\gamma$ quantum lattice fluctuations that were neglected by ME. These fluctuations give rise to a non-zero isotope effect on $m^{\ast}$, which is small unless the system is sufficiently close to a polaronic instability. Within our model, the isotope effect is negative near half filling and positive away from half filling, which is due to the competition of lattice and density fluctuations.

The paper is organized as follows. In Sec. \ref{mome} we introduce the electron-phonon lattice Hamiltonian and its associated effective phonon action within the local DMF formalism and recall the main ideas underlying the adiabatic expansion. In Sec. \ref{apps} we calculate physical quantities ``beyond ME'': the Luttinger-Ward functional, electron self-energy, and isotope effect on electron effective mass. In Sec. \ref{disc} we compare our results to recent experimental findings and draw some general conclusions.
\section{Model and Method} \label{mome}
We study a general tight-binding based Hamiltonian $H = H_{\rm el} + H_{\rm ph} + H_{\rm el-ph}$, where
\begin{eqnarray}
H_{\rm el} & = & -\sum_{ij\sigma}t_{i-j}
(c_{i\sigma}^{\dagger} c_{j\sigma} + c_{j\sigma}^{\dagger} c_{i\sigma}) - N \mu n + H_{\rm ee},
\label{hel} \\
H_{\rm ph} & = & \frac{1}{2} \sum_{i}(M\dot{x}_{i}^2 + Kx_{i}^2),\label{hph} \\  
H_{\rm el-ph} & = & g \sum_{i} x_{i} (n_i - n). \label{helph}
\end{eqnarray}
The operator $c_{i\sigma}^{\dagger}$ creates an electron with spin $\sigma$ on lattice site $i$.  The mean density $n = (1/N)\sum_{i\sigma} c_{i \sigma}^{\dagger} c_{i \sigma}$ (where $N$ is the number of lattice sites) is fixed by adjusting the chemical potential $\mu$. Electron-electron interactions $H_{\rm ee}$ are not explicitly written. The operator $x_i$ measures the ionic displacement at site $i$. From spring constant $K$ and ion mass $M$ we can define the characteristic oscillator frequency $\omega_0 = (K/M)^{1/2}$, which we take to be dispersionless (Einstein model). The electron-phonon interaction couples the phonon displacement $x_i$ to the electron density $n_{i} = \sum_{\sigma} c_{i\sigma}^{\dagger} c_{i\sigma}$ on the same site. We define $\langle x_i \rangle = 0$ to be the equilibrium phonon displacement for a uniform electron distribution.

In DMF theory the properties of $H$ may be obtained from the solution of an impurity model\cite{fjs93} specified by the action $S[c, \bar{c}, x, a] = S_0[x] + S_{\rm ee}[c, \bar{c}, a] + S_1[c, \bar{c}, x, a]$, with 
\begin{eqnarray}
S_0[x] & = & \frac{1}{2T} \sum_{k} x_{k} \left(K + M\omega_k^2 \right) x_{-k}, \label{s0} \\
S_1[c, \bar{c}, x, a] & = & - \sum_{n\,\sigma} \bar{c}_{n \sigma} c_{n \sigma} a_n + g \sum_{n k \sigma} \bar{c}_{n \sigma} c_{n + k,\sigma} x_k. \label{s1}
\end{eqnarray}
The impurity electron is represented by Grassmann fields $c_{n\sigma}, \bar{c}_{n\sigma}$, which depend on odd Matsubara frequencies $\omega_n = (2n+1)\pi T$ and spin $\sigma = \uparrow,\downarrow$. The impurity phonon is represented by a bosonic field $x_k$, which depends on even Matsubara frequencies $\omega_k = 2k\pi T$. Electron-electron interactions (arising from $H_{\rm ee}$) are described by $S_{\rm ee}$. The local electron Green function is defined by $G_{\rm loc}[a]_n = \delta \ln Z[a]/\delta a_n$ where
\begin{equation}
Z[a] = \int [dc d\bar{c} dx] \exp -S[c, \bar{c}, x, a] \label{za}
\end{equation}
is the partition function. The mean field function $a$ (which contains information about the nonlocal physics) is fixed by the self-consistency condition
\begin{equation}
G_{\rm loc}[a]_n = \int d\epsilon_k \frac{\rho(\epsilon_k)}{i\omega_n + \mu -
\Sigma[a]_n - \epsilon_k}, \label{gscc}
\end{equation}
which equates $G_{\rm loc}[a]$ and the momentum integrated lattice Green function. The momentum integral has been converted to an energy integral using the lattice density of states $\rho(\epsilon_k)$.

We integrate out the electron fields and work with an effective phonon action $S[x, a] = S_0[x] +S_{\rm ee}[a] + S_1[x, a] = -\ln \int [dc d\bar{c}] \exp(-S[c,\bar{c},x,a])$. In practice $S$ may depend on additional auxiliary fields (such as spin and charge fluctuation fields), which have to be averaged over. We then use the crucial fact\cite{m58} that the scale $\omega_0 = (K/M)^{1/2}$ on which the phonon fields $x$ vary is much smaller than the scale $t$ (bandwidth or interaction scale) on which electronic quantities (such as $G$ and $a$) vary so that an expansion is possible in the ``adiabatic parameter'' $\gamma = \omega_0/t \ll 1$. In a first step we formally expand $S_1$ in powers of $x$ about $\bar{x} = 0$ (corresponding to a conventional metallic state with no lattice distortions), using\cite{dm00} $x^n \sim \gamma^{n/2}$. In a second step [to be performed below, in the analysis leading up to Eqs. (\ref{s1an})--(\ref{s2bn})] the phonon vertices of the action may be evaluated via a low-frequency expansion.
\section{Physical quantities} \label{apps}
\begin{figure}
\vspace{1cm}
\includegraphics[width=5cm]{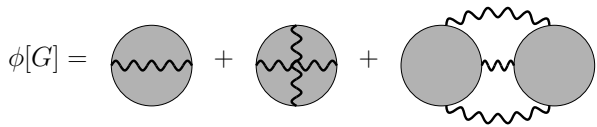}
\vspace{1cm}
\caption{Luttinger-Ward functional to ${\cal O}(\gamma^2)$. A wavy line stands for the phonon Green function $D$. The shaded circles stand for the phonon vertices $\tilde{\Gamma}_2$, $\tilde{\Gamma}_3$, and $\tilde{\Gamma}_4$.}
\label{LW}
\end{figure}

In the following we will not directly work with the effective phonon action but with the Luttinger-Ward functional $\phi^{\rm ph}$ derived from it. In general $\phi = \phi^{\rm ee} + \phi^{\rm ph}$ is defined as the sum of all vacuum-to-vacuum skeleton diagrams. Within the DMF approximation it is related to the local free energy $\Omega_{\rm imp}$ via\cite{gkkr96}
\begin{equation}
\beta \Omega_{\rm imp} = \phi[G] + \sum_{n \sigma} \left( \ln G_{n\sigma} - \Sigma_{n\sigma} G_{n\sigma} \right). \label{omimp}
\end{equation}
The electron self-energy $\Sigma = \Sigma^{\rm ee} + \Sigma^{\rm ph}$ is a saddle point of $\Omega_{\rm imp}$. This implies $\Sigma = \delta \phi/\delta G$. Since the thermodynamic potential $\Omega_{\rm imp}$ [given by Eq. (\ref{omimp})] and the partition function $Z$ [given by Eq. (\ref{za})] are related via $\exp (-\beta \Omega_{\rm imp}) = Z$ we may obtain $\phi^{\rm ph}$ from an adiabatic expansion of $Z$. To order $\gamma^2$ we find
\begin{eqnarray}
&& \phi^{\rm ph}[G] = \frac{1}{2} \sum_k \ln D_k^{-1} \nonumber \\
&& - \frac{\lambda^2 T}{4t}\sum_{k k^{\prime}} \tilde{\Gamma}_4[G]_{k,k^{\prime},-k,-k^{\prime}} D_k  D_{k^{\prime}} \nonumber \\
&& - \frac{\lambda^3 T}{3t}\sum_{k k^{\prime}} \tilde{\Gamma}_3[G]^2_{k,k^{\prime},k-k^{\prime}} D_k D_{k^{\prime}} D_{k-k^{\prime}}, \label{phiG}
\end{eqnarray}
where $D_k = (1 + (\omega_k/\omega_0)^2 - \lambda \tilde{\Gamma}_2[G]_{k, -k})^{-1}$ is the phonon Green function and $\tilde{\Gamma}_n[G]_{k_1,\ldots,k_n}$ are dimensionless phonon vertices whose explicit form depends on $S_{\rm ee}$. The phonon fields $gx \rightarrow x$ were rescaled to have units of energy. The parameter $\lambda = g^2/(Kt)$ is the electron-phonon interaction strength. The Feynman diagrams corresponding to Eq. (\ref{phiG}) are shown in Fig. \ref{LW}.

We use Eq. (\ref{phiG}) to formally calculate $\Sigma$ to ${\cal O}(\gamma^2)$. Following the ``adiabatic expansion'' outlined above we may further simplify the resulting expressions by evaluating the $\tilde{\Gamma}_n$ (whose dominant contributions come from frequencies of order $t$) via a low-frequency expansion. To the given order in $\gamma$ we may approximate the cubic and quartic vertices by their static value (which we write $\tilde{\Gamma}_3$ and $\tilde{\Gamma}_4$) and the phonon Green function by $D_k \approx (1 - \lambda/\lambda_c + (\omega_k/\omega_0)^2 + \lambda \gamma \alpha_p |\omega_k/\omega_0| + {\cal O}(\gamma^2))^{-1}$, where $\lambda_c = \tilde{\Gamma}_2[G]^{-1}_{0,0}$ is a critical interaction strength and $\alpha_p$ is a damping parameter. The static $\tilde{\Gamma}_2$ vertex acts as a phonon self-energy and renormalizes the expansion parameters to $\bar{\gamma} = \gamma (1 - \lambda/\lambda_c)^{1/2}$ and $\bar{\lambda} = \lambda/(1 - \lambda/\lambda_c)$. If $\lambda \rightarrow \lambda_c$ from below then $\bar{\lambda} \rightarrow \infty$. This ``polaronic instability'' was analyzed in great detail in Ref. \onlinecite{dm00}. In the following we assume $\lambda < \lambda_c$. The self-energy to ${\cal O}(\gamma^2)$ can be written as $\Sigma^{\rm ph} = \Sigma_{1a} + \Sigma_{1b} + \Sigma_{1c} + \Sigma_{2a} + \Sigma_{2b}$ (see Fig. \ref{s2l}), where
\begin{eqnarray}
\Sigma_{1a\,n} & = & \lambda t T \sum_k D_k G_{n + k} \label{s1an}, \\
\Sigma_{1b\,n} & = & \lambda^3 \tilde{\Gamma}_4 T^2 \sum_{k k^{\prime}} D_k^2 D_{k^{\prime}} G_{n + k} \label{s1bn}, \\
\Sigma_{1c\,n} & = & 2\lambda^4 \tilde{\Gamma}_3^2 T^2 \sum_{k k^{\prime}} D_k^2 D_{k^{\prime}} D_{k-k^{\prime}} G_{n + k} \label{s1cn}, \\
\Sigma_{2a\,n} & = & \lambda^2 t^2 T^2 \sum_{k k^{\prime}} D_k D_{k^{\prime}} G_{n + k} G_{n + k^{\prime}} G_{n + k + k^{\prime}}, \label{s2an} \\
\Sigma_{2b\,n} & = & 2\lambda^3 t \tilde{\Gamma}_3 T^2 \sum_{k k^{\prime}} D_k D_{k^{\prime}} D_{k - k^{\prime}} G_{n + k} G_{n + k^{\prime}}. \label{s2bn}
\end{eqnarray}
The ${\cal O}(\gamma)$ one-loop diagram $\Sigma_{1a}$ is at the basis of ME theory. The ${\cal O}(\gamma^2)$ two-loop diagrams $\Sigma_{1b}$, $\Sigma_{1c}$, $\Sigma_{2a}$, and $\Sigma_{2b}$ represent lowest-order corrections to ME. Apart from $\Sigma_{1c}$ they all arise from vertex corrections of the electron-phonon coupling. The diagrams $\Sigma_{1b}$, $\Sigma_{1c}$, and $\Sigma_{2b}$ were overlooked in Refs. \onlinecite{gcp98} and \onlinecite{bz98}.

We now turn to the electron mass enhancement from interactions with phonons, defined by $m^{\ast}/m|_{\rm ph} = 1 - \partial \Sigma^{\rm ph}(i\nu)/\partial (i\nu)|_{\nu = 0} = 1 + \lambda_{1a} + \ldots + \lambda_{2b}$, with $\Sigma^{\rm ph}(i\nu)$ written at $T = 0$. We will calculate $m^{\ast}/m$ to ${\cal O}(\gamma)$. Since the frequency derivative is of ${\cal O}(\gamma^{-1})$ we need $\Sigma$ to ${\cal O}(\gamma^2)$ as above. We first calculate the mass enhancement $\lambda_{1a}$ from the basic one-loop diagram $\Sigma_{1a}(i\nu) = \lambda t \int d\omega/(2\pi) G(i\nu + i\omega) /(1 - \lambda/\lambda_c + (\omega/\omega_0)^2)$. In general ${\rm Im}\,\Sigma_{1a\,R}(\nu)$ is of order $\bar{\lambda} \bar{\gamma}$ and is zero for $|\nu| \leq \bar{\omega}_0$. From this and Eq. (\ref{gscc}) it follows that $G(i\nu) = G^{\rm inc}(i\nu) + G^{\rm coh}(i\nu)$ where
\begin{equation}
G^{\rm inc}(i\nu) = \frac{1}{2t^2}(i\nu + \mu), \quad G^{\rm coh}(i\nu) = -i\pi\,{\rm sign}(\nu) \rho(\mu) \label{ginc}
\end{equation}
at low frequencies, for $S_{\rm ee} = 0$, and neglecting terms of ${\cal O}(\gamma)$. Weak electron-electron interactions renormalize the scale $t$ in Eqs. (\ref{ginc}), but their general form remains the same. We thus find
\begin{equation}
\lambda_{1a} = \bar{\lambda} t \rho(\mu) - \frac{1}{4}\bar{\lambda} \bar{\gamma}, \label{l1a}
\end{equation}
where the leading term comes from the derivative acting on $G^{\rm coh}$, producing a delta function $\delta(\omega)$, and the ${\cal O}(\gamma)$ term comes from $G^{\rm inc}$. The phonon damping $\sim |\omega|$ in $D$ does not give rise to any ${\cal O}(\gamma)$ corrections to (\ref{l1a}). The two-loop diagrams can be analyzed in a similar fashion. We just give the results:
\begin{eqnarray}
\lambda_{1b} & = & \frac{1}{2} \bar{\lambda}^3 \bar{\gamma} t \rho(\mu) \tilde{\Gamma}_4, \label{l1b} \\
\lambda_{1c} & = & \frac{1}{2} \bar{\lambda}^4 \bar{\gamma} t \rho(\mu) \tilde{\Gamma}_3^2, \label{l1c} \\
\lambda_{2a} & = & -\frac{1}{4} \bar{\lambda}^2 \bar{\gamma} t \rho(\mu) \left(3\pi^2 t^2 \rho(\mu)^2 - 5(\mu/2t)^2 \right), \label{l2a} \\
\lambda_{2b} & = & \bar{\lambda}^3 \bar{\gamma} t \rho(\mu) (\mu/2t) \tilde{\Gamma}_3. \label{l2b}
\end{eqnarray}
In principle our adiabatic expansion is based on the smallness of the parameter $\gamma$ alone, with no assumption being made about $\lambda$. However, ${\cal O}(\gamma^n)$ terms in the expansion are of the form $\bar{\lambda}^m \bar{\gamma}^n \sim (1 - \lambda/\lambda_c)^{n/2 - m}$ with $m \geq n$ and become arbitrarily large if $\lambda \rightarrow \lambda_c$. We therefore need to set the range of parameters $\bar{\lambda}$ and $\bar{\gamma}$ where the expansion is valid. We propose two different criteria. 

First we notice\cite{rr} that at half filling $m^{\ast}/m$ as a function of $\bar{\lambda}$ has a local maximum at $\bar{\lambda}_{\rm max\,1}$ and then goes to $-\infty$ as $\bar{\lambda} \rightarrow \infty$. This unphysical behavior comes from the fact that $\tilde{\Gamma}_3$ vanishes at $\mu = 0$ and that the dominant term $\lambda_{1b} \sim \bar{\lambda}^3 \bar{\gamma} \tilde{\Gamma}_4$ to this order is negative. For $0 < |\mu| \leq \mu_1 \ll 1$, i.e. sufficiently close to half filling, the local maximum at $\bar{\lambda}_{\rm max\,1}$ can still be defined, but now $m^{\ast}/m \rightarrow +\infty$ as $\bar{\lambda} \rightarrow \infty$ because the dominant term $\lambda_{1c} \sim \bar{\lambda}^4 \bar{\gamma} \tilde{\Gamma}_3^2$ is positive. The upper bound $\mu_1$ depends weakly on $\gamma$, as shown in the first two columns of Table \ref{lmax1}. We thus suggest that for given $\gamma$ and $|\mu| \leq \mu_1(\gamma)$ our adiabatic expansion is valid if $\bar{\lambda} \leq \bar{\lambda}_{\rm max\,1}$. In Table \ref{lmax1} we give $\bar{\lambda}_{\rm max\,1}$ and the corresponding unrenormalized quantity $\lambda_{\rm max\,1}$ at half filling ($\mu = 0$), for a few representative values of $\gamma$.

Farther away from half filling, i.e. for $|\mu| > \mu_1$, the local maximum in $m^{\ast}/m$ ceases to exist and we need a different criterion $\bar{\lambda}_{\rm max\,2}$, which we define as follows: for given $\gamma$ we need $\bar{\lambda} \leq \bar{\lambda}_{\rm max\,2}$ small enough so that the ratio of the ${\cal O}(\gamma)$ terms and the ${\cal O}(1)$ term in $m^{\ast}/m$ be less than $p < 1$. We fix the parameter $p$ by setting $\bar{\lambda}_{\rm max\,1} = \bar{\lambda}_{\rm max\,2}$ at $\mu = 0$: the result is $p \approx 0.25$, which is compatible with the ${\cal O}(\gamma)$ terms being a small perturbation. In Table \ref{lmax2} we give $\bar{\lambda}_{\rm max\,2}$ and $\lambda_{\rm max\,2}$ at half filling and at $\mu/(2t) = 0.4$, for some values of $\gamma$.

We now return to the effective mass. In the following we assume $S_{\rm ee} = 0$ and a semicircular density of states $\rho(\epsilon_k) = 1/(2\pi t^2) (4t^2 - \epsilon_k^2)^{1/2} \theta(4t^2 - \epsilon_k^2)$, for which the vertices can be evaluated exactly: $\tilde{\Gamma}_2 = 8[1 - (\mu/2t)^2]^{3/2}/(3\pi)$, $\tilde{\Gamma}_3 = 4(\mu/2t) [1 - (\mu/2t)^2]^{3/2}/\pi$ and $\tilde{\Gamma}_4 = -16[1 - (\mu/2t)^2]^{3/2}[1 - 6(\mu/2t)^2]/(15\pi)$. In Fig. \ref{menh} we plot $m^{\ast}/m$ as a function of $\bar{\lambda}$ for three different values of $\gamma$. At half filling ($\mu = 0$, main graph) we plot up to $\bar{\lambda}_{\rm max\,1}$ (cf. Table \ref{lmax1}). Far away from half filling ($\mu/(2t) = 0.4$, inset) we plot up to $\bar{\lambda}_{\rm max\,2}(0.4)$ (cf. Table \ref{lmax2}).  

The isotope effect on electron effective mass is defined by $\alpha_{m^{\ast}} = -d \ln m^{\ast}/d \ln M$. Using $M \sim \gamma^{-2}$ it is convenient to rewrite this as $\alpha_{m^{\ast}} = \gamma/(2m^{\ast}) dm^{\ast}/d\gamma$, so that
\begin{equation}
\alpha_{m^{\ast}} = \frac{-\frac{1}{4} \bar{\lambda} \bar{\gamma} + \lambda_{1b} + \ldots + \lambda_{2b}}{2(1 + \bar{\lambda} t \rho(\mu))} + {\cal O}(\gamma^2). \label{am}
\end{equation}
In the main graph of Fig. \ref{amst} we plot $\alpha_{m^{\ast}}$ as a function of $\bar{\lambda} \leq \bar{\lambda}_{\rm max\,2}(0)$ (cf. Table \ref{lmax2}), at half filling and for different values of $\gamma$. The isotope effect is negative and its absolute value is limited by $0.125 = p/2$. In the inset of Fig. \ref{amst} we plot $\alpha_{m^{\ast}}$ at $\mu/(2t) = 0.4$, for the same values of $\gamma$. For $1.5 \lesssim \bar{\lambda} \leq \bar{\lambda}_{\rm max\,2}(0.4)$, the isotope effect is positive. We explain this in the next section.
\section{Discussion and conclusion} \label{disc}
\begin{figure}
\vspace{1cm}
\includegraphics[width=5.5cm]{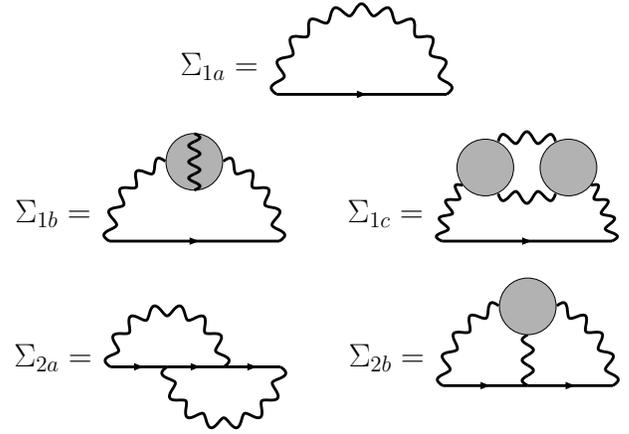}
\vspace{1cm}
\caption{Self-energy diagrams to ${\cal O}(\gamma^2)$.  A straight line stands for $G$, a wavy line stands for $D$. The phonon vertices $\tilde{\Gamma}_3$ and $\tilde{\Gamma}_4$ are represented by shaded circles.}
\label{s2l}
\end{figure}

\begin{figure}
\includegraphics[width=8cm]{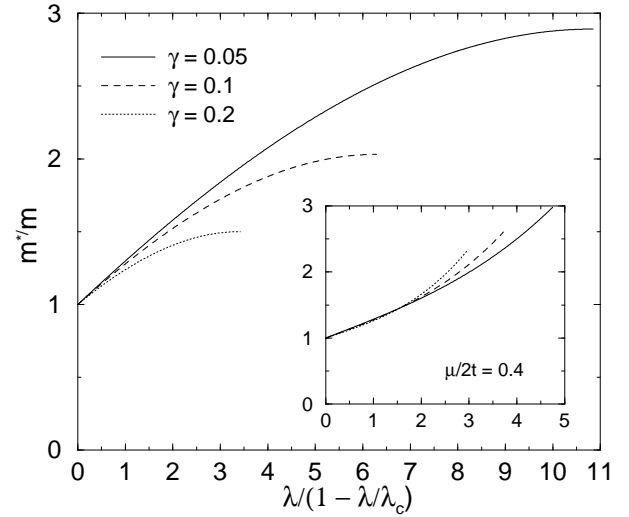}
\caption{Electron mass enhancement $m^{\ast}/m$ due to interactions with phonons, as a function of renormalized electron-phonon coupling $\bar{\lambda}$ and for different values of adiabatic parameter $\gamma = \omega_0/t$. Main graph: half filling ($\mu = 0$) and $\bar{\lambda} \leq \bar{\lambda}_{\rm max\,1}(\gamma)$. Inset: $\mu/(2t) = 0.4$ and $\bar{\lambda} \leq \bar{\lambda}_{\rm max\,2}(\gamma)$.}
\label{menh} 
\end{figure}

\begin{figure}
\includegraphics[width=8cm]{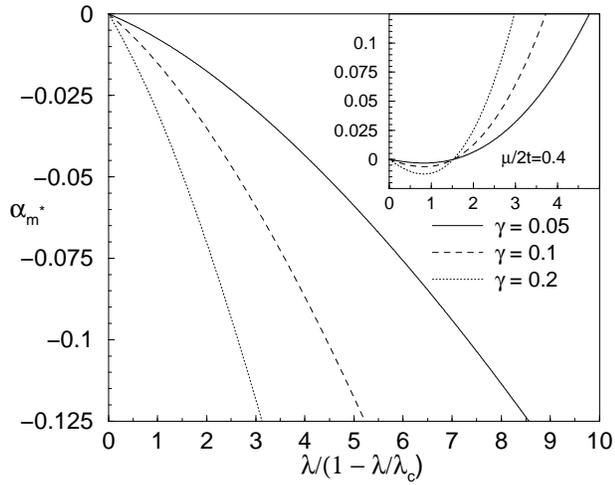}
\caption{Effective mass isotope exponent $\alpha_{m^{\ast}} = -d \ln
m^{\ast}/$\\$d \ln M$ as a function of renormalized electron-phonon coupling $\bar{\lambda}$ for different values of adiabatic parameter $\gamma = \omega_0/t$, at $T = 0$ and within the range of validity of the adiabatic expansion. Main graph: half filling ($\mu = 0$). Inset: away from half filling ($\mu/(2t) = 0.4$).}
\label{amst}
\end{figure}

We start with a brief analysis of $\Sigma$. The correct two-loop self-energy consists of four diagrams, three of which, the diagrams containing the vertices $\tilde{\Gamma}_3$ and $\tilde{\Gamma}_4$, were neglected by previous authors.\cite{gcp98,bz98} Fig. \ref{s2l} and Eqs. (\ref{s1bn}), (\ref{s1cn}) suggest that the diagrams $\Sigma_{1b}$ and $\Sigma_{1c}$ can be absorbed in a $\gamma$-dependent $\lambda_c$ as follows: $\lambda_{c}^{-1}(\gamma) = \tilde{\Gamma}_2 + (A_1\,\bar{\lambda}\,\tilde{\Gamma}_4 + B_1\,\bar{\lambda}^2\,\tilde{\Gamma}_3^2 )\,\bar{\gamma} + {\cal O}(\bar{\gamma}^2)$. If we assume $T \ll \omega_0$ (which is appropriate for calculating $m^{\ast}$) then $A_1$ and $B_1$ (which arise from the summation over $D_k$) are positive constants of order one. As pointed out in the previous section, the ${\cal O}(\gamma)$ term in $\lambda_c(\gamma)$ changes sign when the system is doped away from half filling. The same is true for the mass enhancement $\lambda_{2a} + \lambda_{2b}$ from the remaining two diagrams. We explain this as follows. Very close to half filling, low-frequency lattice fluctuations reduce the average local ion displacement $x$. As a result, the electron-phonon coupling in Eq. (\ref{helph}) is less efficient, $m^{\ast}/m$ decreases relative to the value $1 + \lambda_{1a}$ corresponding to a completely static lattice, and the isotope effect is negative. This is the physics behind the terms $\lambda_{1b}$ and $\lambda_{2a}$ and the second part of $\lambda_{1a}$. It is reflected in the main graphs of Figs. \ref{menh} and \ref{amst}. Away from half filling, on the other hand, the physics is dominated by density fluctuations, which have the opposite effect. They enhance the density-coupled interaction (\ref{helph}), increase $m^{\ast}/m$, and lead to a positive $\alpha_{m^{\ast}}$. This is represented by the terms $\lambda_{1c}$ and $\lambda_{2b}$ and in the insets of Figs. \ref{menh} and \ref{amst}.

We now comment on the subject of expansion parameters, which has been the source of some controversy in the literature. The original ME articles\cite{m58} and more recent work\cite{gps95} indicate that $\lambda \gamma$ is the proper expansion parameter of a nonadiabatic theory, whereas other authors (see, e.g., Ref. \onlinecite{mms96} and references therein and, for a particularly clear discussion, Ref. \onlinecite{bz98}) argue that the expansion breaks down if $\lambda$ exceeds a critical value of order one, irrespective of $\gamma$. Within our method, it is possible to reconcile these two points of view. By working with renormalized parameters $\bar{\lambda}$ and $\bar{\gamma}$ we implicitly take into account the presence of the polaronic instability at $\lambda_c$. The condition $\lambda < \lambda_c$ of Refs. \onlinecite{bz98} and \onlinecite{mms96} thus appears as a natural limitation of our theory. If $\lambda$ is sufficiently close to the instability at $\lambda_c$ (which is possible even for $\bar{\lambda} \leq \bar{\lambda}_{\rm max}(\gamma)$, as can be seen from Tables \ref{lmax1} and \ref{lmax2}) then the dominant mass enhancement terms $\lambda_{1b} \sim (1 - \lambda/\lambda_c)^{-5/2}$ (close to half filling) and $\lambda_{1c} \sim (1 - \lambda/\lambda_c)^{-7/2}$ (away from half filling) act by shifting $\lambda_c$ to higher and lower values, respectively, as shown in the previous paragraph. This argument can immediately be extended to higher orders in $\gamma$, noting that the odd vertices $\tilde{\Gamma}_{2n+1}$ all vanish at half filling due to partcle-hole symmetry. We thus conclude that the effective expansion parameters of our theory are $\bar{\lambda} \bar{\gamma}$ close to half filling and $\bar{\lambda}^2 \bar{\gamma}$ away from half filling. This generalizes the results of Refs. \onlinecite{m58} and \onlinecite{gps95}.

Next we compare our calculations to experiments. Ref. \onlinecite{zhkm97} presented measurements of the oxygen isotope dependence of the Meissner fraction of LSCO high-temperature superconductors. The assumption that the dependence arose from an isotope effect on the carrier mass $m^{\ast\ast}$ (and hence the penetration depth) implied $\alpha_{m^{\ast\ast}} \approx -0.5$. Within our theory, such large values of $\alpha_{m^{\ast\ast}}$ imply that the material is at or beyond the polaronic instability; however, this assumption may not be consistent with the observed reasonably good conductance of high-$T_c$ materials. On the other hand, Refs. \onlinecite{fbpfzgm01} and \onlinecite{sbflzzg01}  have demonstrated that in the ``colossal magnetoresistance'' material ${\rm La}_{1-x}{\rm Ca}_x {\rm MnO}_3$ ($0.2 \leq x \leq 0.5$) the isotope effect on low-temperature properties including the carrier mass (determined via specific heat) is very small, even though electron-lattice interactions are believed to be strong in this material and even though the isotope effect on the ferromagnetic transition temperature $T_C$ is large.\cite{zckm96,bbgkbok98,ficci98,heil00} In the manganites it seems likely that the very large isotope effects on $T_C$ are associated with a phase transition that is now believed to be first order;\cite{almas00} we speculate that a similar phenomenon might explain the data on the high-$T_c$ material.

In summary, we have presented a detailed theory of isotope effects in models with strong electron-lattice coupling. We have determined the correct expansion parameters and have found diagrams overlooked in previous works. Our results suggest that a large isotope effect on electronic properties is very difficult to obtain in a metallic system. Unlike previous treatments of nonadiabatic effects,\cite{gcp98,gps95} our formalism can easily be generalized to include electron correlations, which enter in two places: (i) vertex and self-energy corrections renormalize the phonon vertices $\tilde{\Gamma}_n$ and (ii) irreducible particle-hole scattering vertices renormalize the electron-phonon coupling $g$. Both effects are believed to strongly suppress the electron-phonon interaction. An extensive analysis of problem (i) for $\tilde{\Gamma}_2$ in the presence of local electron-electron interactions can be found in Ref. \onlinecite{dm00}. Progress on (ii) will be presented in a future publication.

We thank S.  Blawid and R.  L.  Greene for useful discussions.  We acknowledge NSF DMR00081075 and the University of Maryland-Rutgers MRSEC for support.

\begin{table}
\caption{Limiting values of $\bar{\lambda}$ and $\lambda$ as a function of $\gamma$ at half filling, using the first criterion (valid for $|\mu| \leq \mu_1$) described in the text.}
\begin{tabular}{llll}
$\gamma$ & $\mu_1$ & $\bar{\lambda}_{\rm max\,1}(0)$ & $\lambda_{\rm max\,1}(0)$ \\
\tableline
0.05 & 0.062 & 10.8407 & 1.0626 \\
0.1  & 0.079 & 6.2804 & 0.9920  \\
0.15 & 0.091 & 4.4458 & 0.9313  \\
0.2  & 0.100 & 3.4168 & 0.8760  \\
0.25 & 0.109 & 2.7456 & 0.8244 
\end{tabular} \label{lmax1}
\end{table}

\begin{table}
\caption{Limiting values of $\bar{\lambda}$ and $\lambda$ as a function of $\gamma$ at half filling and at $\mu/(2t) = 0.4$, using the second criterion described in the text.}
\begin{tabular}{lllll}
$\gamma$ & $\bar{\lambda}_{\rm max\,2}(0)$ & $\lambda_{\rm max\,2}(0)$ & $\bar{\lambda}_{\rm max\,2}(0.4)$ & $\lambda_{\rm max\,2}(0.4)$ \\
\tableline
0.05 & 8.5620 & 1.0356 & 4.7589 & 1.1579 \\
0.1  & 5.2225 & 0.9613 & 3.7243 & 1.0846 \\
0.15 & 3.8793 & 0.9037 & 3.2578 & 1.0412 \\
0.2  & 3.1268 & 0.8557 & 2.9782 & 1.0109 \\
0.25 & 2.6369 & 0.8143 & 2.7874 & 0.9879
\end{tabular} \label{lmax2}
\end{table}
\end{multicols}
\end{document}